\documentclass[aps,preprint,showpacs,showkeys]{revtex4}
\usepackage{amsmath}
\usepackage{bm}

\input{tcilatex}

\begin{document}

\title{Calculation of the eigenfunctions and eigenvalues of Schr\"{o}dinger
type equations by asymptotic Taylor expansion method (ATEM)}
\author{Ramazan Ko\c{c}}
\altaffiliation{Corresponding Author: E-mail address: koc@gantep.edu.tr, Phone: +90 342 317
2209, Fax: +90 342 360 18 22}
\author{ Eser Ol\u{g}ar}
\altaffiliation{E-mail address: olgar@gantep.edu.tr}
\affiliation{Gaziantep University, Department of Physics, Faculty of Engineering 27310
Gaziantep/Turkey}
\date{\today }

\begin{abstract}
A novel method is proposed to determine an analytical expression for
eigenfunctions and numerical result for eigenvalues of the Schr\"{o}dinger
type equations, within the context of Taylor expansion of a function.
Optimal truncation of the Taylor series gives a best possible analytical
expression for eigenfunctions and numerical result for eigenvalues.
\end{abstract}

\maketitle

\section{Introduction}

One of the source of progress of the sciences depends on the study of the
same problem from different point of view. Besides their progress of the
sciences those different point of views include a lot of mathematical
tastes. Over the years considerable attention has been paid to the solution
of Schr\"{o}dinger equation. Determination of eigenvalues of the Schr\"{o}%
dinger equation via asymptotic iteration method (AIM) has recently attracted
some interest, arising from the development of fast computers \cite%
{ciftci1,ciftci2,ciftci3,saad}. This method have been widely applied to
establish eigenvalues of the Schr\"{o}dinger type equations \cite%
{koc1,koc2,barakat,fernandez,soylu,olgar1}. Although the AIM formalism is
very efficient to obtain eigenvalues of the Schr\"{o}dinger equation, it
requires tedious calculations in order to determine wave function if the
system is not exactly solvable. When Schr\"{o}dinger equation includes a non
solvable potential, the calculation of wave function involved with a large
number of terms will lose its simplicity and accuracy.

In this paper we will discuss new formalism based on the Taylor series
expansion method, namely Asymptotic Taylor Expansion Method (ATEM).\
Although, Taylor Series Method \cite{taylor} is an old one, it appears,
however, not has been fully exploited in the analysis of both in solution of
physical and mathematical problems. Yet, even today, new contributions to
this problem are being made \cite{hu}. Apart from its formal relation to
AIM, ATEM has also been easily applied to solve second order linear
differential equations by introducing a simple computer program. We would
like to mention here that the ATEM is a field of tremendous scope and has an
almost unlimited opportunity, for its applications in the solution of the
Schr\"{o}dinger type equations. One can display a number of fruitful
applications of the ATEM in different fields of the physics. For instance,
our formalism of ATEM gives a new approach to the series solution of the
differential equations as well as interrelation between series solution of
differential equations and AIM. The method can also be applied to solve
Dirac equation and Klein Gordon equation. In this paper we address ourselves
to the solution of the eigenvalue problems by using the ATEM.

One of the fundamental advantage of ATEM is that (approximate) analytical
expressions for the wave function of the associated Hamiltonian can easily
be obtained. We note that ATEM\ also gives an accurate result for the
eigenvalues when it is compared to AIM.

It should also be noted here that the determination of wave function by
using AIM require tedious calculations if the system is not exactly
solvable. For non exactly solvable potentials, the calculation of wave
function by using AIM involved with a large number of terms will lose its
simplicity and accuracy. Therefore, the method introduced here is useful to
determine an analytical expression for the wave function of the non exactly
solvable equations.

The paper is organized as follows. The first main result of the paper is
given in section by reformulating the well known Taylor series expansion of
a function. Section 3 is devoted to the application of the main result for
solving the Schr\"{o}dinger equation including various potentials. As a
practical example, we illustrate solution of the Schr\"{o}dinger equation
including anharmonic oscillator potential and the Hamiltonian of an
interacting electron in a quantum dot. In this section we present an
approximate analytical expression for eigenfunction and numerical results
for eigenvalues of the anharmonic oscillator potential and the Hamiltonian
of an interacting electron in a quantum dot. We also analyze asymptotic
behavior of the Hamiltonian. Finally we comment on the validity of our
method and remark on its possible use in different fields of the physics in
section 4.

\section{A new formalism of the Taylor Expansion Method}

In this section, we show the solution of the Schr\"{o}dinger equation for a
quite ample class of potentials, by modifying Taylor series expansion by
means of a finite sequence instead of an infinite sequence and its
termination possessing the property of quantum mechanical wave function. In
quantum mechanics bound state energy of the atom is quantized and
eigenvalues are discrete and for each eigenvalues there exist one or more an
eigenfunctions. When we are dealing with the solution of the Schr\"{o}dinger
equation we are mainly interested in the discrete eigenvalues of the
problem. The first main result of this conclusion gives necessary and
sufficient conditions for the termination of the Taylor series expansion of
the wave function.

Let us consider Taylor series expansion of a function $f(x)$ about the point 
$a$:%
\begin{eqnarray}
f(x) &=&f(a)+(x-a)f^{\prime }(a)+\frac{1}{2}(x-a)^{2}f^{\prime \prime }(a)+%
\frac{1}{6}(x-a)^{3}f^{(3)}(a)+\cdots  \notag \\
&&\overset{\infty }{=\underset{n=0}{\dsum }}\frac{(x-a)^{n}}{n!}f^{(n)}(a)
\label{a1}
\end{eqnarray}%
where $f^{(n)}(a)$ is the $n^{th}$ derivative of the function at $a$. Taylor
series specifies the value of a function at one point, $x$, in terms of the
value of the function and its derivatives at a reference point $a$.
Expansion of the function $f(x)$ about the origin ($a=0$), is known as
Maclaurin's series and it is given by,%
\begin{eqnarray}
f(x) &=&f(0)+xf^{\prime }(0)+\frac{1}{2}x^{2}f^{\prime \prime }(0)+\frac{1}{6%
}x^{3}f^{(3)}(0)+\cdots  \notag \\
&&\overset{\infty }{=\underset{n=0}{\dsum }}\frac{x^{n}}{n!}f^{(n)}(0).
\label{a11}
\end{eqnarray}%
Here we develop a method to solve a second order linear differential
equation of the form: 
\begin{equation}
f^{\prime \prime }(x)=p_{0}(x)f^{\prime }(x)+q_{0}(x)f(x).  \label{a2}
\end{equation}%
It is obvious that the higher order derivatives of the $f(x)$ can be
obtained in terms of the $f(x)$ and $f^{\prime }(x)$ by differentiating (\ref%
{a2}). Then, higher order derivatives of $f(x)$ are given by%
\begin{equation}
f^{(n+2)}(x)=p_{n}(x)f^{\prime }(x)+q_{n}(x)f(x)  \label{a3}
\end{equation}%
where%
\begin{eqnarray}
p_{n}(x) &=&p_{0}(x)p_{n-1}(x)+p_{n-1}^{\prime }(x)+q_{n-1}(x),\text{ and} 
\notag \\
q_{n}(x) &=&q_{0}(x)p_{n-1}(x)+q_{n-1}^{\prime }(x).  \label{a4}
\end{eqnarray}%
Of course, the last result shows there exist a formal relation between AIM
and ATEM. To this end, we conclude that the recurrence relations (\ref{a4})
allow us algebraic exact or approximate analytical solution of (\ref{a2})
under some certain conditions. Let us substitute (\ref{a4}) into the (\ref%
{a1}) to obtain the function that is related to the wave function of the
corresponding Hamiltonian:%
\begin{equation}
f(x)=f(0)\left( 1+\overset{m}{\underset{n=2}{\dsum }}q_{n-2}(0)\frac{x^{n}}{%
n!}\right) +f^{\prime }(0)\left( 1+\overset{m}{\underset{n=2}{\dsum }}%
p_{n-2}(0)\frac{x^{n}}{n!}\right) .  \label{a5}
\end{equation}%
After all we have obtained an useful formalism of the Taylor expansion
method. This form of the Taylor series can also be used to obtain series
solution of the second order differential equations. In the solution of the
eigenvalue problems, truncation of the the asymptotic expansion to a finite
number of terms is useful. If the series optimally truncated at the smallest
term then the asymptotic expansion of series is known as superasymptotic 
\cite{boyd}, and it leads to the determination of eigenvalues with minimum
error.

Arrangement of the boundary conditions for different problems becomes very
important because improper sets of boundary conditions may produce
nonphysical results. When only odd or even power of $x$ collected as
coefficients of $f(0)$ or $f^{\prime }(0)$ and vice verse, the series is
truncated at $n=m$ then an immediate practical consequence of these
condition for $q_{m-2}(0)=0$ or $p_{m-2}(0)=0.$ In this way, the series
truncates at $n=m$ and one of the parameter in the $q_{m-2}(0)$ or $%
p_{m-2}(0)$ belongs to the spectrum of the Schr\"{o}dinger equation.
Therefore eigenfunction of the equation becomes a polynomial of degree $m$.
Otherwise the spectrum of the system can be obtained as follows: In a
quantum mechanical system eigenfunction of the system is discrete. Therefore
in order to terminate the eigenfunction $f(x)$ we can concisely write that%
\begin{eqnarray}
q_{m}(0)f(0)+p_{m}(0)f^{\prime }(0) &=&0  \notag \\
q_{m-1}(0)f(0)+p_{m-1}(0)f^{\prime }(0) &=&0  \label{a5y}
\end{eqnarray}%
eliminating $f(0)$ and $f^{\prime }(0)$ we obtain%
\begin{equation}
q_{m}(0)p_{m-1}(0)-p_{m}(0)q_{m-1}(0)=0  \label{a5z}
\end{equation}%
again one of the parameter in the equation related to the eigenvalues of the
problem.

We can state that the ATEM reproduces exact solutions to many exactly
solvable differential equations and these equations can be related to the
Schr\"{o}dinger equation. It will be shown in the following section ATEM
also gives accurate results for non-solvable Schr\"{o}dinger equations, such
as the sextic oscillator, cubic oscillator, deformed Coulomb potential, etc.
which are important in applications to many problems in physics. This
asymptotic approach opens the way to the treatment of \ Schr\"{o}dinger type
equation including large class of potentials of practical interest.

\section{Solution of the Schr\"{o}dinger equation by using ATEM}

An analytical solution of the Schr\"{o}dinger equation is of high importance
in nonrelativistic quantum mechanics, because the wave function contains all
necessary information for full description of a quantum system. In this
section we take a new look at the solution of the Schr\"{o}dinger equation
by using the method of ATEM developed in the previous section. Let us
consider the following eigenvalue problem ($\hbar =2m=1$)%
\begin{equation}
-\frac{d^{2}\psi (x)}{dx^{2}}+V(x)\psi (x)=E\psi (x)  \label{a51}
\end{equation}%
where $V(x)$ is the potential, $\psi (x)$ is wave function and $E$ is the
energy of the system. The equation has been solved exactly for a large
number of potentials by employing various techniques. In general, it is
difficult to determine the asymptotic behavior of (\ref{a51}) in the present
form. Therefore it is worthwhile to transform (\ref{a51}) to an appropriate
form by introducing the wave function $\psi (x)=f(x)\exp \left( -\dint
W(x)dx\right) $. Thus, this change of wave function guaranties $\underset{%
x->\infty }{\lim }\psi (x)=0$. We recast (\ref{a51}) and we obtain the
following equation%
\begin{equation}
L(x)=-f^{\prime \prime }(x)+2W(x)f^{\prime }(x)+(V(x)+W^{\prime
}(x)-W^{2}(x)-E)f(x)=0.  \label{a61}
\end{equation}%
In this formalism of the equation coefficients in (\ref{a2}) can be
expressed as:%
\begin{equation*}
p_{0}(x)=2W(x),\text{\qquad }q_{0}(x)=(V(x)+W^{\prime }(x)-W^{2}(x)-E).
\end{equation*}%
Using the relation given in (\ref{a4}) one can easily compute $p_{n}(x)$ and 
$q_{n}(x)$ by a simple MATHEMATICA program. Our task is now to illustrate
the use of ATEM to obtain explicit analytical solution of the Schr\"{o}%
dinger equation including various potentials.

\subsubsection{Anharmonic oscillator}

Solution of the Schr\"{o}dinger equation including anharmonic potential has
attracted a lot of attention, arising its considerable impact on the various
branches of physics as well as biology and chemistry. The equation is
described by the Hamiltonian%
\begin{equation}
H=-\frac{d^{2}}{dx^{2}}+x^{2}+gx^{4}.  \label{o1}
\end{equation}%
In practice anharmonic oscillator problem is always used to test accuracy
and efficiency of the unperturbative methods. Let us introduce, the
asymptotic solutions of anharmonic oscillator Hamiltonian when $W(x)=x$,
then the wave function takes the form%
\begin{equation*}
\psi =e^{-\frac{x^{2}}{2}}f(x),
\end{equation*}%
and (\ref{a61} )can be expressed as%
\begin{equation}
L(x)=-\frac{d^{2}f}{dx^{2}}+2x\frac{df}{dx}+(gx^{4}+1-E)f=0.  \label{a71}
\end{equation}%
Comparing the equations (\ref{a2}) and (\ref{a71}) we can deduce that%
\begin{equation}
p_{0}(x)=2x\ \text{and }q_{0}(x)=(gx^{4}+1-E).  \label{a72}
\end{equation}%
Here we take a new look at the solution of the (\ref{a71}) by using the
method of ATEM developed in the previous section. By applying (\ref{a5z}),
the corresponding energy eigenvalues are calculated by the aid of a
MATHEMATICA program.

The term asymptotic means the function approaching to a given value as the
iteration number tends to infinity. By the aid of MATHEMATICA program we
calculate eigenvalues $E$and eigenfunction $f(x)$ for $g=0.1$ using number
of iterations $k=\{20,30,40,50,60,70,80\}$. The eigenvalues are presented in
Table I and and are compared with results computed by the AIM \cite{ciftci1}
and direct numerical integration method \cite{must} by taking $g=0.1.$

\begin{table}[tbph]
\begin{tabular}{|l|l|l|l|l|l|l|}
\hline
$k\ \ \ \ \ \ \ $ & $n=0\ \ \quad \qquad $ & $n=1\qquad $ & $\qquad n=2$ & $%
n=3$ & $n=4$ & $n=5$ \\ \hline
$20$ & $1.06529019$ & $3.30632658$ & $5.74394288$ & $8.30136568$ & $%
10.82233628$ & $15.77123716$ \\ \hline
$30$ & $1.06528554$ & $3.30688248$ & $5.74807461$ & $8.35361636$ & $%
11.10902943$ & $13.84616319$ \\ \hline
$40$ & $1.06528550$ & $3.30687176$ & $5.74795553$ & $8.35266513$ & $%
11.09828874$ & $13.97716619$ \\ \hline
$50$ & $1.06528550$ & $3.30687202$ & $5.74795940$ & $8.35267765$ & $%
11.09860503$ & $13.96951294$ \\ \hline
$60$ & $1.06528550$ & $3.30687201$ & $5.74795926$ & $8.35267786$ & $%
11.09859535$ & $13.96995159$ \\ \hline
$70$ & $1.06528550$ & $3.30687201$ & $5.74795926$ & $8.35267782$ & $%
11.09859562$ & $13.96992450$ \\ \hline
$80$ & $1.06528550$ & $3.30687201$ & $5.74795926$ & $8.35267782$ & $%
11.09859562$ & $13.96992632$ \\ \hline
$E$ \cite{ciftci1} & $1.065286$ & $3.306871$ & $5.747960$ & $8.352642$ & $%
11.09835$ & $13.96695$ \\ \hline
$E$ \cite{must} & $1.065286$ & $3.306872$ & $5.747959$ & $8.352678$ & $%
11.09860$ & $13.96993$ \\ \hline
\end{tabular}%
\caption[The comparison of eigenvalues]{Eigenvalues of the (\protect\ref{o1}%
) for different iteration numbers $k$ and $g=0.1.$Last two rows corresponds
the comparison of eigenvalues computed by the AIM \protect\cite{ciftci1},
direct numerical integration method \protect\cite{must} .}
\end{table}
The function $f(x)$ for $n=2$ state is given in (\ref{eq1}).

\begin{eqnarray}
k=30;f(x)= &&1-2.37404x^{2}+0.147996x^{4}+0.0193758x^{6}  \notag \\
&&-1.73022\times 10^{-3}x^{8}-5.18734\times 10^{-5}x^{10}+8.68491\times
10^{-6}x^{12}  \notag \\
k=50;f(x)= &&1-2.37398x^{2}+0.14797x^{4}+0.0193735x^{6}  \notag \\
&&-1.73037\times 10^{-3}x^{8}-5.19242\times 10^{-5}x^{10}+8.67726\times
10^{-6}x^{12}  \label{eq1} \\
k=80;f(x)= &&1-2.37398x^{2}+0.14797x^{4}+0.0193735x^{6}  \notag \\
&&-1.73037\times 10^{-3}x^{8}-5.19243\times 10^{-5}x^{10}+8.67725\times
10^{-6}x^{12}  \notag
\end{eqnarray}

As we mentioned before using ATEM we can obtain an analytical expression for
the wave function of the Schr\"{o}dinger equation. Substituting $E$ into (%
\ref{a5}) we get the wave function of the Schr\"{o}dinger equation for the
corresponding eigenvalues. Analytical expressions signal that ATEM produce
an efficient result for the eigenfunction. For the first four states the
plot of the normalized wave functions are given in Figure 1.

\begin{figure}[tbp]
\caption{The first four wavefunctions for anharmonic oscillator.}
\QTP{Dialog Text}
\begin{tabular}{|l|l|}
\hline
\FRAME{itbpF}{3.3797in}{2.3255in}{0in}{}{}{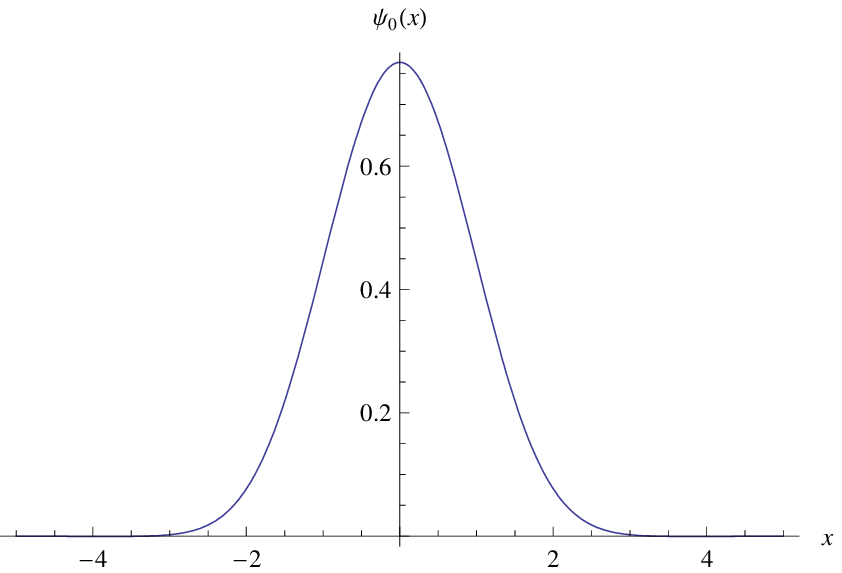}{\special{language
"Scientific Word";type "GRAPHIC";maintain-aspect-ratio TRUE;display
"USEDEF";valid_file "F";width 3.3797in;height 2.3255in;depth
0in;original-width 3.333in;original-height 2.2857in;cropleft "0";croptop
"1";cropright "1";cropbottom "0";filename 'fig0.eps';file-properties
"XNPEU";}} & \FRAME{itbpF}{3.3797in}{2.2157in}{0in}{}{}{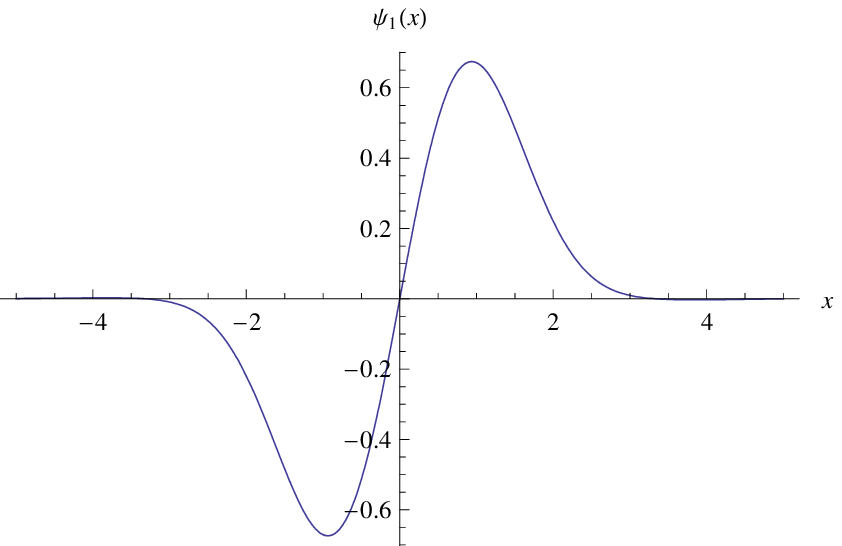}{\special%
{language "Scientific Word";type "GRAPHIC";maintain-aspect-ratio
TRUE;display "USEDEF";valid_file "F";width 3.3797in;height 2.2157in;depth
0in;original-width 3.333in;original-height 2.175in;cropleft "0";croptop
"1";cropright "1";cropbottom "0";filename 'fig1.eps';file-properties
"XNPEU";}} \\ \hline
\FRAME{itbpF}{3.3797in}{2.2157in}{0in}{}{}{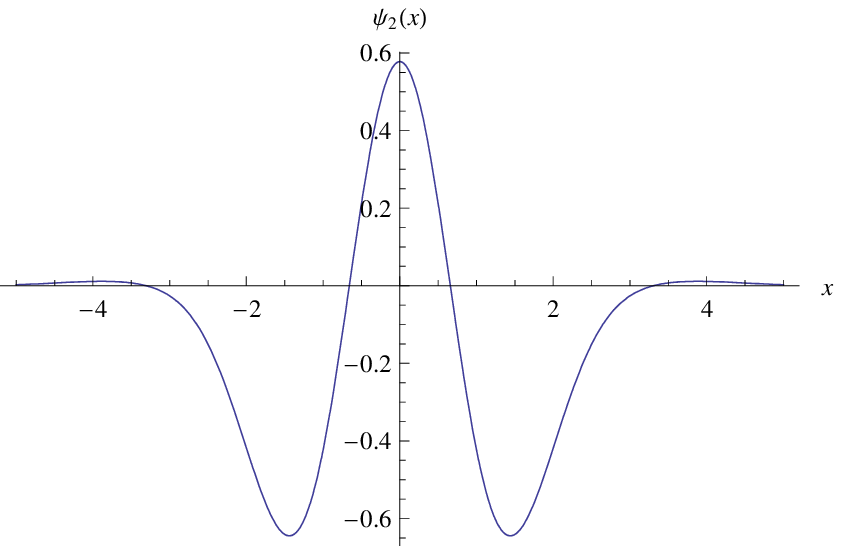}{\special{language
"Scientific Word";type "GRAPHIC";maintain-aspect-ratio TRUE;display
"USEDEF";valid_file "F";width 3.3797in;height 2.2157in;depth
0in;original-width 3.333in;original-height 2.175in;cropleft "0";croptop
"1";cropright "1";cropbottom "0";filename 'fig2.eps';file-properties
"XNPEU";}} & \FRAME{itbpF}{3.3797in}{2.2157in}{0in}{}{}{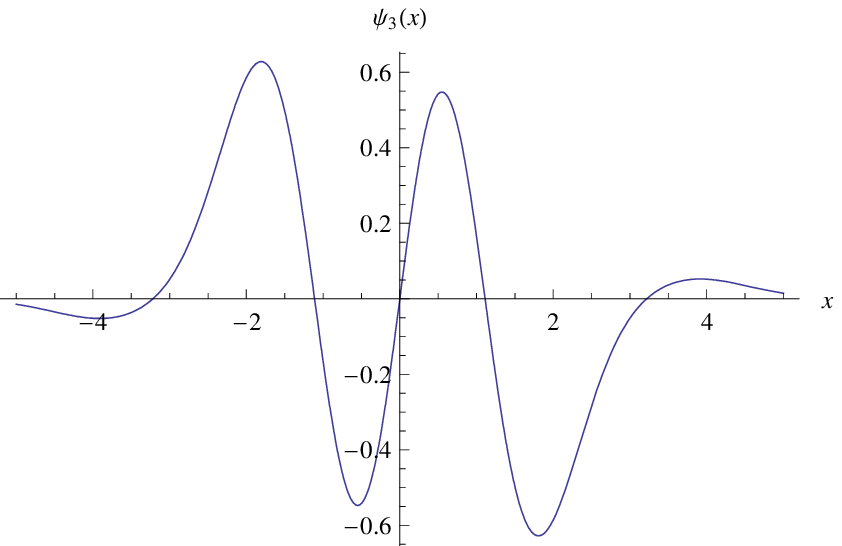}{\special%
{language "Scientific Word";type "GRAPHIC";maintain-aspect-ratio
TRUE;display "USEDEF";valid_file "F";width 3.3797in;height 2.2157in;depth
0in;original-width 3.333in;original-height 2.175in;cropleft "0";croptop
"1";cropright "1";cropbottom "0";filename 'fig3.eps';file-properties
"XNPEU";}} \\ \hline
\end{tabular}%
\end{figure}

\subsubsection{Interacting electrons in a quantum dot}

In this section we present a procedure to solve the Schr\"{o}dinger equation
of two interacting electrons in a quantum dot in the presence of an external
magnetic field by using ATEM. The problem has been discussed in various
articles \cite{olgar2,turbiner,taut}. Here we just solve the mathematical
part of the problem. Without further discussion the Schr\"{o}dinger equation
for a quantum dot containing two electrons in the presence of the magnetic
field $B$ perpendicular to the dot is given by%
\begin{equation}
H=\sum\limits_{i=1}^{2}(\frac{1}{2m_{i}^{\ast }}(P_{i}+eA(r_{i}))^{2}+\frac{1%
}{2}m_{i}^{\ast }\omega _{0}^{2}r_{i}^{2})+\frac{e^{2}}{\varepsilon
\left\vert r_{2}-r_{1}\right\vert }  \label{e4}
\end{equation}%
Introducing relative and center of mass coordinates $r=r_{2}-r_{1},\quad R=%
\frac{1}{2}(r_{1}+r_{2})$ the Hamiltonian can be separated into two parts
such that $H=2H_{r}+\frac{1}{2}H_{R},$ where 
\begin{subequations}
\begin{eqnarray}
H_{r} &=&\frac{p^{2}}{2m^{\ast }}+\frac{1}{2}m^{\ast }\omega ^{2}r^{2}+\frac{%
e^{2}}{2\epsilon r}+\frac{1}{2}\omega _{c}L_{r},  \label{e5a} \\
H_{R} &=&\frac{P^{2}}{2m^{\ast }}+\frac{1}{2}m^{\ast }\omega ^{2}R^{2}+\frac{%
1}{2}\omega _{c}L_{R}.  \label{e5b}
\end{eqnarray}%
Equation (\ref{e5b}) is the Hamiltonian of the harmonic oscillator, and it
can be solved exactly. Let us turn our attention to the solution of the
Hamiltonian $H_{r}$. In the polar coordinate $r=(r,\alpha )$, if the
eigenfunction 
\end{subequations}
\begin{equation}
\phi =r^{-\frac{1}{2}}e^{i\ell \alpha }u(r)  \label{e7}
\end{equation}%
is introduced, the Schr\"{o}dinger equation $H_{r}\phi =E_{r}\phi $, can be
expressed as%
\begin{equation}
\left( -\frac{\hbar ^{2}}{2m^{\ast }}\frac{d^{2}}{dr^{2}}+\frac{\hbar ^{2}}{%
2m^{\ast }}(\ell ^{2}-\frac{1}{4})\frac{1}{r^{2}}+\frac{1}{2}m^{\ast }\omega
^{2}r^{2}+\frac{e^{2}}{2\epsilon r}+\frac{1}{2}\omega _{c}L_{r}\right)
u(r)=E_{r}u(r).  \label{e8}
\end{equation}%
From now on we restrict ourselves to the solution of \ Eq. (\ref{e8}). After
changing the variable $r\rightarrow \frac{\hbar }{\sqrt{2m^{\ast }}}r$ and
substituting $u(r)=r^{\ell +\frac{1}{2}}e^{-\frac{\hbar \omega }{4}%
r^{2}}f(r),$we obtain the following equation%
\begin{equation}
L(r)=-rf^{\prime \prime }(r)+(\omega r^{2}-(2\ell +1))f^{\prime
}(r)-(rE_{n}+\lambda )f(r)=0,  \label{a62}
\end{equation}%
where%
\begin{equation}
E_{r}=E_{n}+(\left\vert \ell \right\vert +1)\hbar \omega +\frac{1}{2}%
L_{r}\omega _{c},\quad \lambda =-\frac{e^{2}}{2\varepsilon },  \label{e14}
\end{equation}%
for simplicity we have chosen that $\hbar \omega =\frac{\hbar ^{2}}{2m^{\ast
}}=1.$ In this case the functions $p_{0}(r)$ and $q_{0}(r)$ are given by 
\begin{equation}
p_{0}(r)=\omega r-\frac{2\ell +1}{r}\ \text{and }q_{0}(r)=-E_{n}-\frac{%
\lambda }{r}.  \label{e15}
\end{equation}%
Meanwhile we bring to mind that Hamiltonian (\ref{e4}) possesses a hidden
symmetry. This implies that the Hamiltonian is quasi-exactly solvable \cite%
{olgar2,turbiner}. Fortunately, quasi exact solvability of the Hamiltonian
gives us an opportunity to check accuracy of our result and to test our
method. In order to obtain quasi exact solution of (\ref{a62}) we set in:%
\begin{equation*}
E_{n}=j\omega ,\text{ where }j=1,2,3....
\end{equation*}%
and then the problem is exactly solvable when the following relation is
satisfied:%
\begin{eqnarray*}
\lambda &=&\left\{ \pm \sqrt{\omega (2\ell +1)}\right\} ;j=1 \\
\lambda &=&\left\{ 0,\pm \sqrt{2\omega (4\ell +3)}\right\} ;j=2 \\
\lambda &=&\left\{ 0,\pm \sqrt{10\omega (\ell +1)\pm \omega \sqrt{73+128\ell
+64\ell ^{2}}}\right\} ;j=3 \\
&&\cdots .
\end{eqnarray*}%
Note that $\lambda ,\omega $ and $\ell $ belong to the spectrum of the
Hamiltonian. Therefore an accuracy check for the ATEM can be made. We have
tested ATEM and the result are given by%
\begin{eqnarray*}
\lambda &=&\left\{ \pm \sqrt{\omega (2\ell +1)}\right\} ;j=1;E_{n}=\omega ;
\\
f(r) &=&1\mp \sqrt{\frac{\omega }{2\ell +1}}r \\
\lambda &=&\left\{ \pm \sqrt{2\omega (4\ell +3)}\right\} ;j=2;E_{n}=2\omega ;
\\
f(r) &=&1\mp \sqrt{\frac{2\omega (4\ell +3)}{2\ell +1}}r+\frac{\omega }{%
2\ell +1}r^{2}.
\end{eqnarray*}%
Consequently, we demonstrated that our approach is able to reproduce exact
results for the exactly solvable second order differential equations. Let us
turn our attention to the complete solution of the (\ref{a62}). We have
again used $80$ iterations during the solution of the equation and
controlled the stability of the eigenvalues. The results are given in Table
II.

\begin{table}[tbph]
\begin{tabular}{|l|l|l|l|l|l|l|}
\hline
$k\ \ \ \ \ \ \ $ & $n=0\quad \qquad $ & $n=1\qquad $ & $\qquad n=2$ & $n=3$
& $n=4$ & $n=5$ \\ \hline
$20$ & $0.63844692$ & $0.81078941$ & $2.43413075$ & $2.80439505$ & $%
4.30282729$ & $4.81597264$ \\ \hline
$30$ & $0.64590375$ & $0.80463157$ & $2.45078087$ & $2.79146457$ & $%
4.32486753$ & $4.79696796$ \\ \hline
$40$ & $0.65007015$ & $0.80121630$ & $2.45998729$ & $2.78436605$ & $%
4.33697615$ & $4.78670663$ \\ \hline
$50$ & $0.65279850$ & $0.79898546$ & $2.46598524$ & $2.77974956$ & $%
4.34484756$ & $4.78008225$ \\ \hline
$60$ & $0.65475424$ & $0.79738773$ & $2.47027173$ & $2.77645083$ & $%
4.35046813$ & $4.77536829$ \\ \hline
$70$ & $0.65624076$ & $0.79617357$ & $2.47352334$ & $2.77394739$ & $%
4.35473042$ & $4.77180017$ \\ \hline
$80$ & $0.65741800$ & $0.79521187$ & $2.47609490$ & $2.77196620$ & $%
4.35810109$ & $4.76898147$ \\ \hline
\end{tabular}%
\caption[The comparison of eigenvalues]{Eigenvalues $E_{n}$ of the (\protect
\ref{a62}) for different iteration numbers $k$ and $\protect\lambda =\protect%
\sqrt{\protect\omega }=1$ and $\ell =1/2.$}
\end{table}
We go back (\ref{a5}) to obtain the wave function of the equation of(\ref%
{a62}) for various values of $E.$ Their plots are given in Figure 2.

\begin{figure}[tbp]
\caption{The wavefunctions of the two electron interacting in the harmonic
oscillator potential field. The parameters $\protect\lambda =\protect\sqrt{%
\protect\omega }=1.$}
\QTP{Dialog Text}
\begin{tabular}{ll}
\FRAME{itbpF}{3.3797in}{2.2157in}{0in}{}{}{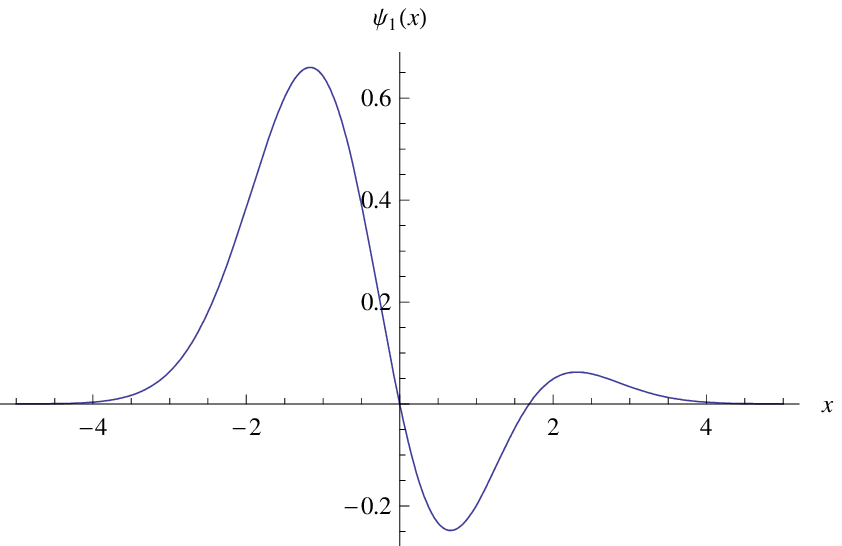}{\special{language
"Scientific Word";type "GRAPHIC";maintain-aspect-ratio TRUE;display
"USEDEF";valid_file "F";width 3.3797in;height 2.2157in;depth
0in;original-width 3.333in;original-height 2.175in;cropleft "0";croptop
"1";cropright "1";cropbottom "0";filename 'fig1c.eps';file-properties
"XNPEU";}} & \FRAME{itbpF}{3.3797in}{2.2157in}{0in}{}{}{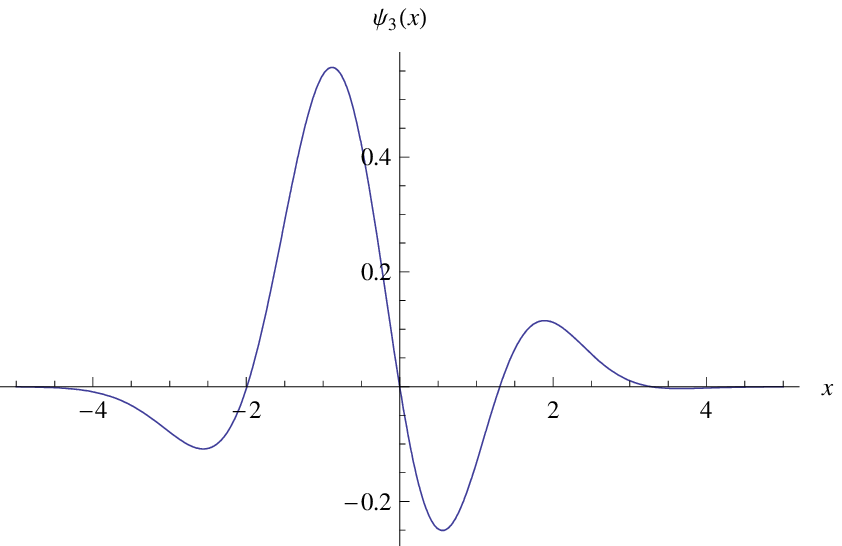}{\special%
{language "Scientific Word";type "GRAPHIC";maintain-aspect-ratio
TRUE;display "USEDEF";valid_file "F";width 3.3797in;height 2.2157in;depth
0in;original-width 3.333in;original-height 2.175in;cropleft "0";croptop
"1";cropright "1";cropbottom "0";filename 'fig2c.eps';file-properties
"XNPEU";}} \\ 
\FRAME{itbpF}{3.3797in}{2.2157in}{0in}{}{}{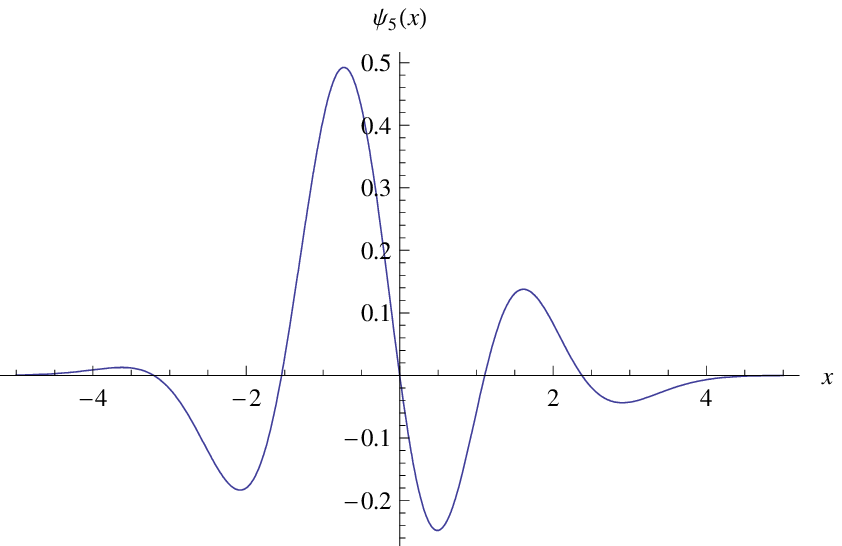}{\special{language
"Scientific Word";type "GRAPHIC";maintain-aspect-ratio TRUE;display
"USEDEF";valid_file "F";width 3.3797in;height 2.2157in;depth
0in;original-width 3.333in;original-height 2.175in;cropleft "0";croptop
"1";cropright "1";cropbottom "0";filename 'fig3c.eps';file-properties
"XNPEU";}} & \FRAME{itbpF}{3.3797in}{2.2157in}{0in}{}{}{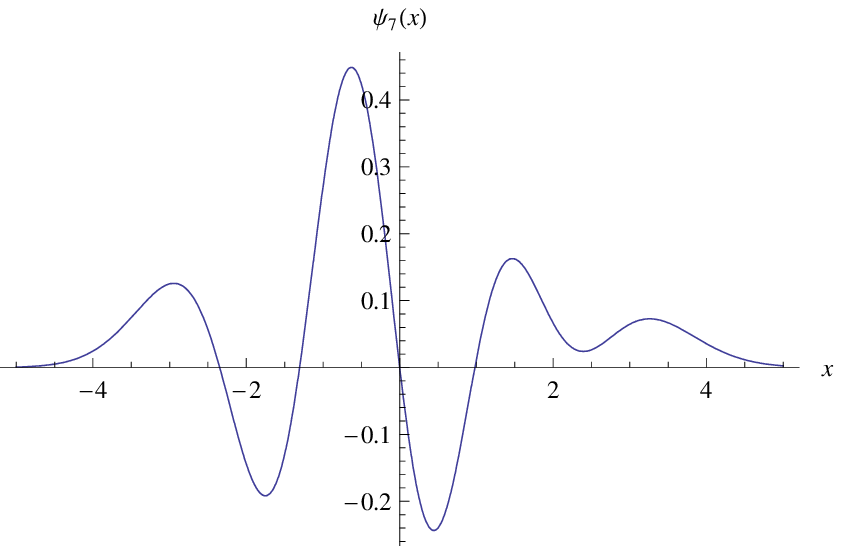}{\special%
{language "Scientific Word";type "GRAPHIC";maintain-aspect-ratio
TRUE;display "USEDEF";valid_file "F";width 3.3797in;height 2.2157in;depth
0in;original-width 3.333in;original-height 2.175in;cropleft "0";croptop
"1";cropright "1";cropbottom "0";filename 'fig4c.eps';file-properties
"XNPEU";}}%
\end{tabular}%
\end{figure}

\section{Conclusion}

The basic features of our approach are to reformulate Taylor series
expansion of a function for obtaining both eigenvalues and eigenfunctions of
the Schr\"{o}dinger type equations. Furthermore the technique given here has
been applied to determine the eigenvalues and the eigenfunctions of the
anharmonic oscillator and the Hamiltonian of two electrons in a quantum dot.
We have shown that ATEM gives accurate results for eigenvalue problems.

As a further work ATEM can be developed in various directions. Position
dependent mass Hamiltonians \cite{koca1,koca2,koc3} can be solved by
extending the method given in this paper. In particular, Lie algebraic or
bosonic Hamiltonians can be solved within the framework of the method given
here. Before ending this work a remark is in order. This extension leads to
the solution of various matrix Hamiltonians, Dirac equation and Klein-Gordon
equation. Our present results manifest that ATEM leads to the solution of
the Schr\"{o}dinger type equations in different fields of physics.

\section{Acknowledgement}

The research was supported by the Scientific and Technological Research
Council of \ TURKEY (T\"{U}B\.{I}TAK).

\end{document}